\documentclass[aps,prb,twocolumn,groupedaddress]{revtex4-2}

\usepackage{graphicx}
\usepackage{dcolumn}
\usepackage{bm}

\usepackage{xcolor}
\usepackage{amssymb}
\usepackage{wasysym}

\newcommand{\RbZn}{$\theta$-(ET)$_2$RbZn(SCN)$_4$}
\newcommand{\TlZn}{$\theta$-(ET)$_2$TlZn(SCN)$_4$}

\newcommand{\thRbZn}{$\theta_o$-RbZn}
\newcommand{\thTlZn}{$\theta_m$-TlZn}

\begin{document}

\title{Comparison of the charge-crystal and charge-glass state in geometrically frustrated organic conductors studied by fluctuation spectroscopy}

\author{Tatjana Thomas}
\author{Tim Thyzel}
\author{Hungwei Sun}
\affiliation{Institute of Physics, Goethe University Frankfurt, 60438 Frankfurt (M), Germany}

\author{Kenichiro Hashimoto}
\affiliation{Department of Advanced Materials Science, University of Tokyo, 277-8561 Chiba, Japan}
\affiliation{Institute for Materials Research, Tohoku University, 980-8577 Sendai, Japan}
\author{Takahiko Sasaki}
\affiliation{Institute for Materials Research, Tohoku University, 980-8577 Sendai, Japan}

\author{Hiroshi M. Yamamoto}
\affiliation{Institute for Molecular Science, Okazaki, 444-8585 Aichi, Japan}

\author{Jens M\"uller}
\email[Email: ]{j.mueller@physik.uni-frankfurt.de}
\affiliation{Institute of Physics, Goethe University Frankfurt, 60438 Frankfurt (M), Germany}

\date{\today}

\begin{abstract}

We present a systematic investigation of the low-frequency charge carrier dynamics in different charge states of the organic conductors $\theta$-(BEDT-TTF)$_2$$M$Zn(SCN)$_4$ with $M$=Rb,Tl, which result from quenching or relaxing the charge degrees of freedom on a geometrically frustrated triangular lattice. Due to strong electronic correlations these materials exhibit a charge-ordering transition, which can be kinetically avoided by rapid cooling resulting in a so-called charge-glass state without long-range order. The combination of fluctuation spectroscopy and a heat pulse method allows us to study and compare the resistance fluctuations in the low-resistive quenched and the high-resistive charge-ordered state, revealing striking differences in the respective noise magnitudes. For both compounds, we find strongly enhanced resistance fluctuations right at the metal-insulator transition and a broad noise maximum in the slowly cooled charge-crystal state with partly dominating two-level processes revealing characteristic activation energies.

\end{abstract}

\maketitle

\section{Introduction}

Recently, the formation of non-equilibrium metastable states when quenching charge degrees of freedom in condensed matter has attracted great interest \cite{Kagawa2017}. Such a new state has been discussed in the organic conductors $\theta$-(BEDT-TTF)$_2$$MM^\prime$(SCN)$_4$, where BEDT-TTF represents bis-ethylenedithio-tetrathiafulvalene (in short: ET) and $MM^\prime$(SCN)$_4$ a monovalent anion with $M=$(Rb,Cs,Tl) and $M^\prime=$(Co,Zn). These quasi-two-dimensional charge-transfer salts consist of an alternating structure of conducting layers with the [(ET)$_2$]$^+$ molecules separated by insulating anion layers [$MM^\prime$(SCN)$_4$]$^-$, where the charge transfer of two donor ET molecules to one acceptor molecule results in a quarter-filled (hole) conduction band. Due to strong electronic correlations the systems often exhibit a charge-ordered (CO) ground state, which can be described within the framework of the extended Hubbard model \cite{Hubbard1963} under consideration of the nearest-neighbor lattice site interaction $V$ in addition to the on-site Coulomb repulsion $U$ \cite{Seo2006}. However, it was found that the first-order CO transition can be kinetically avoided by rapid cooling leading to a so-called charge-glass state which lacks long-range order \cite{Kagawa2013}. This state had been interpreted in purely electronic terms by considering the frustration of charges due to the geometric arrangement of the ET molecules on an anisotropic triangular lattice within the conducting layer. Therefore, a suitable parameter to characterize the degree of frustration is the ratio of the inter-site Coulomb interactions $V_2/V_1$ along different crystallographic axes (see Appendix \ref{Crystal_structure}), which can be altered either by pressure or by chemical variations of the anion atoms $M$ and $M^\prime$ \cite{Mori1998}. This parameter has a severe influence on the critical cooling rate $|q_{\text{c}}|$ required to avoid charge ordering, where for the compounds with orthorhombic crystal structure a clear systematics is found, in which a stronger frustration leads to a higher charge-glass forming ability and therefore to lower critical cooling rates \cite{Sato2014a,Kagawa2017}. In the monoclinic compound \TlZn, in short \thTlZn, the large anisotropy of the triangular lattice and thus low geometric frustration results in a high critical cooling rate $|q_{\text{c}}|>50\,$K/min \cite{Sasaki2017}, whereas in the orthorhombic system \RbZn, in short \thRbZn, with medium frustration a charge-glass state is reported for cooling rates $|q_{\text{c}}|\gtrsim5\,$K/min \cite{Kagawa2013}. The two compounds exhibit different CO patterns, i.e. a diagonal pattern for \thTlZn\ and a horizontal one for \thRbZn, which results from different strengths of electron-phonon coupling \cite{Udagawa2007,Miyashita2007}, where the monoclinic \thTlZn\ approximates a system where the effects are purely electronic in nature \cite{Sasaki2017}.\\
Fluctuation (noise) spectroscopy has proven to be a powerful tool to study the low-frequency dynamics of the charge carriers in molecular metals providing information on inhomogeneous current distributions, slow dynamics due to a glassy freezing of structural or electronic degrees of freedom, or the slowing down of the charge carrier kinetics in the vicinity of a critical endpoint, see e.g.\ \cite{JMueller2011,JMueller2018} and references therein. Indeed, for the title compounds $\theta$-(ET)$_2$$MM^\prime$(SCN)$_4$, measurements of the resistance noise above the metal-insulator transition in \thRbZn, $\theta_o$-CsZn and \thTlZn\ revealed slow and heterogeneous dynamics \cite{Kagawa2013,Sato2014,Sato2016,Sasaki2017}, which are ascribed to vitrification/glassy freezing of the dynamics of charge clusters upon approaching the glass-transition temperature from above. Most remarkably, the process of charge crystallization in the supercooled state was found to be very similar to that of conventional glass-forming liquids. Studying charges on a geometrically frustrated lattice in molecular metals can therefore help to better understand the physics of glasses in general \cite{Sasaki2017,Sato2017}. Contrary to the interpretation of the charge-glass formation in purely electronic terms, recent thermal expansion combined with resistance fluctuation spectroscopy measurements on the highly frustrated $\theta_o$-CsZn and $\theta_o$-CsCo have provided evidence for a \textit{structural} glassy transition at $T_{\text{g}}\sim90-100\,$K \cite{Thomas2022}. This and strong structural changes at the metal-insulator transition in \thRbZn\ \cite{Mori1998, Watanabe2004, Alemany2015} raise the question, to what extent structural degrees of freedom are involved in the development of different charge states in the less geometrically frustrated compounds and highlights the need for further investigations of the novel charge-glass state.\\
In this work, we extend the previous studies of the charge carrier dynamics, which were restricted to temperatures above the metal-to-insulator transition, to a systematic investigation of all three different charge states, namely the charge-liquid (CL) state at high temperatures $T > T_{\rm CO}$, where the charge is homogeneously distributed in space, the ordered charge-crystal (CC) state for $T < T_{\rm CO}$ for slow cooling and the metastable charge-glass (CG) state, when CO is kinetically avoided by fast cooling. The different states are investigated in orthorhombic \thRbZn\ as well as in the monoclinic variant of \thTlZn, which exhibit a different strength of charge frustration and electron-phonon coupling. For both compounds, we find strongly enhanced and slow resistance fluctuations right at the transition into the charge-ordered phase as well as a broad noise maximum below $T_{\rm CO}$. In \thRbZn, the fluctuations in the charge-crystal state are dominated by two-level processes characterized by slow and heterogeneous dynamics. In addition, deviations from a quadratic current dependence of the power spectral density of current fluctuations are observed, which coincide with the onset of nonlinear $IV$ curves in the charge-ordered state, whereas for \thTlZn\ a quadratic scaling of the noise was observed. In both systems, the comparison of the charge carrier dynamics in the quenched CG state reveals a much lower noise magnitude than that of the CC state. 

\section{Experiment}

\begin{figure}[b]
	\centering
	\includegraphics[width=0.75\linewidth]{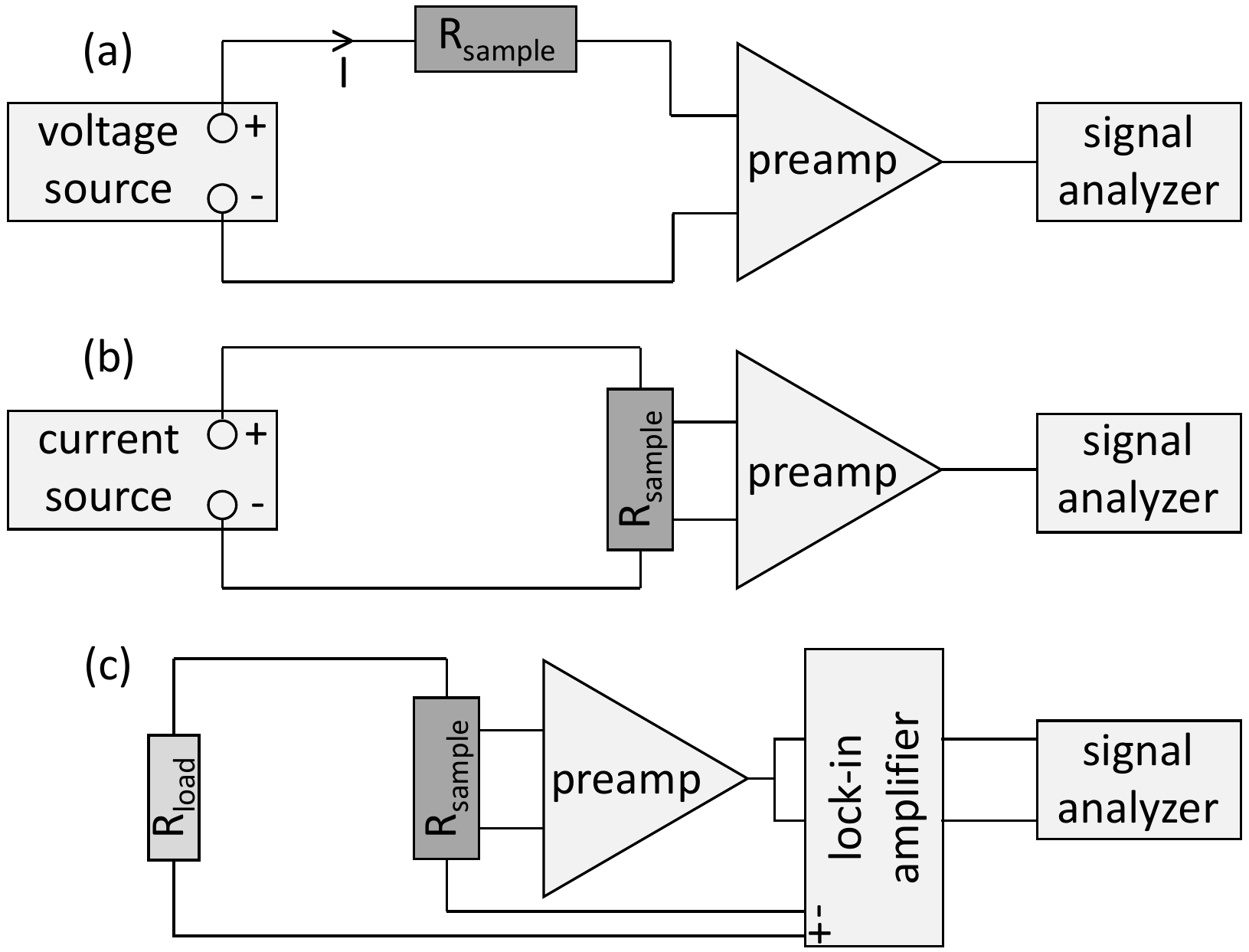}
	\caption{Different setups for fluctuation spectroscopy: (a) Two-point DC setup ($\ocircle$) for measurements of current (conductance) fluctuations in high-impedance samples, (b) four-point DC setup ($\square$) and (c) four-point AC setup ($\triangle$) for measurements of voltage (resistance) fluctuations in more conductive samples. The fluctuating signal in each case is fed into a signal analyzer (SR785) to obtain the noise power spectral density (PSD). The data symbols shown in the figures below indicate which setup has been used.}
	\label{fig:noise_setup}
\end{figure}

Single crystals of $\theta$-(ET)$_2$$MM^\prime$(SCN)$_4$ were grown by electrochemical crystallization \cite{Mori1998}. $\theta$-RbZn crystals were grown with an improved method that does not use 18-crown-6 ether. Rb(SCN) and Zn(SCN)$_2$ were solubilized by using MeOH/1,1,2-Trichloroethane (1:9 v/v) as a solvent in the electrochemical process. Resistance measurements have been performed parallel to the conducting layers in a DC configuration, for which 10 or 25\,$\mu$m-thick gold wires were attached to the crystal using carbon paste.
Due to the large variation of the samples' impedances, sometimes of several orders of magnitude, upon cooling down from room temperature, the measurement setup for fluctuation spectroscopy, schematically shown in Fig.\ \ref{fig:noise_setup}, was varied such that four-terminal AC and DC techniques were used to measure voltage fluctuations and a two-terminal DC conductance method for measuring current fluctuations, see \cite{JMueller2011,JMueller2018} for more detailed information.

\begin{figure}[t]
	\centering
	\includegraphics[width=1\linewidth]{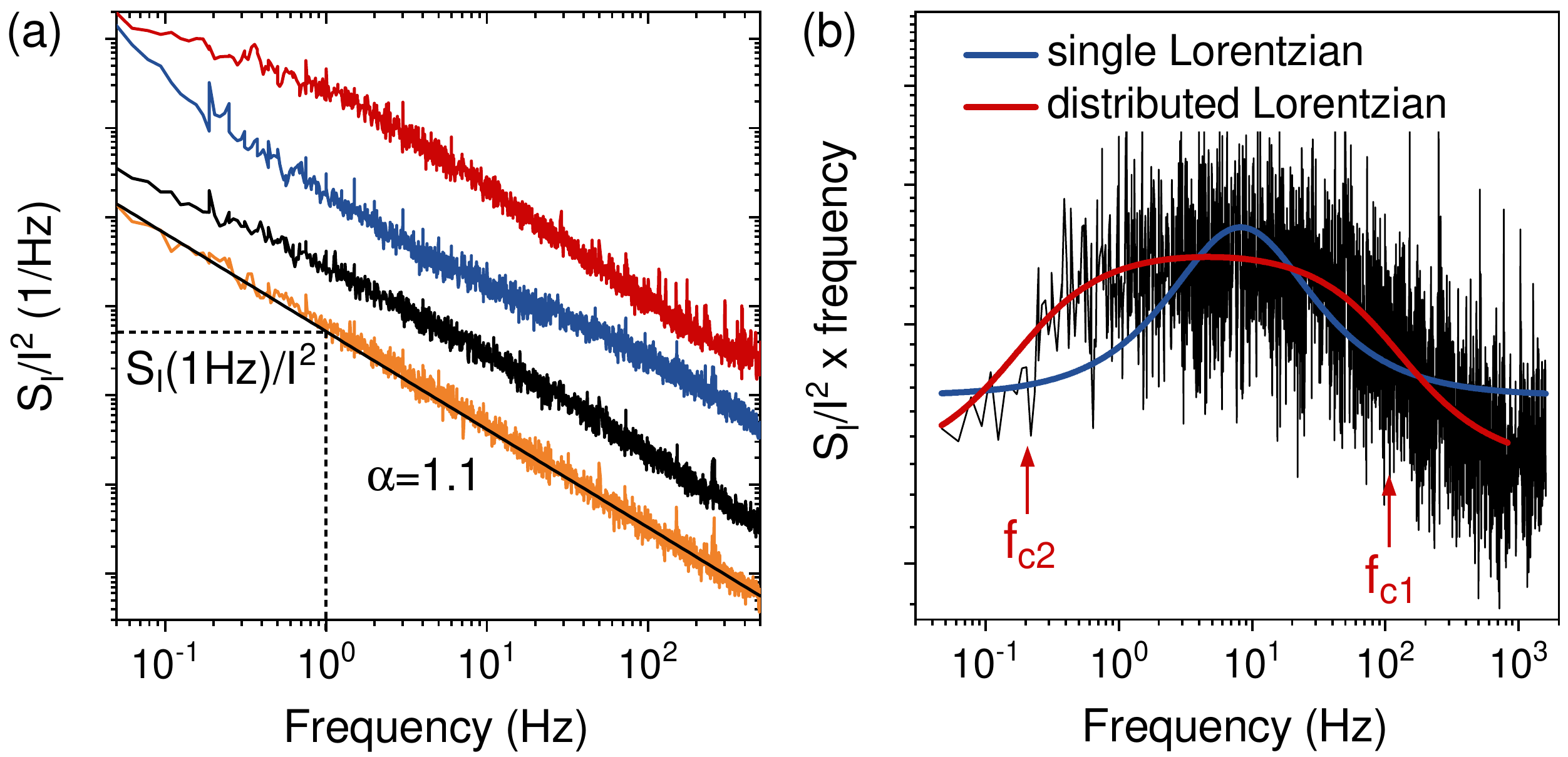}
	\caption{(a) Typical noise spectra either of $1/f$-type or superimposed with a Lorentzian contribution for selected temperatures in \thRbZn\ (sample \#1). (b) Comparison of the single and distributed Lorentzian model, where the latter can better reproduce the line width of the measured spectrum.}
	\label{fig:spectra}
\end{figure}

\noindent
A constant DC voltage or current was supplied by Keithley sourcemeters (K2400 or K2612), whereas a lock-in amplifier (Stanford Research Systems SR830) combined with a load resistor was used for the AC technique. For the two-point conductance method the sample is connected in series with a current amplifier (NF CA5350 or Keithley K428), where the internal resistance determines the gain and transforms the current to a voltage signal. For the four-terminal setups a preamplifier (SR560) was used. The main advantage of the AC technique is the minimization of the premaplifier's own, extrinsic $1/f$ noise contribution by choosing a suitable excitation frequency to operate in the ``eye'' of the  preamplifier's noise figure \cite{Scofield1987}. The fluctuating signal is further processed by a signal analyzer (Stanford Research Systems SR785), which calculates the Fast Fourier Transform and provides the power spectral density (PSD) $S_{V,I}(f)$ of the voltage or current fluctuations. Usually this quantity exhibits a quadratic dependence on the applied voltage or current in the ohmic regime \cite{Hooge1969,Kogan1996} so that the normalized PSD $S_V/V^2$ or $S_I/I^2$, conveniently taken at 1\,Hz, is used for analyzing the temperature-dependent noise magnitude on different samples. It was always ensured that the measured noise originates from the sample and not from the preamplifier or other external sources and that the normalized spectra of different techniques yield the same results, i.e.\ $S_V/V^2 = S_I/I^2 = S_R/R^2 = S_G/G^2$, where the indices denote voltage, current, resistance and conductance, respectively. The observed noise spectra were either of pure $1/f$-type or composed of underlying $1/f$ noise superimposed with a single or distributed Lorentzian contribution, exemplarily shown in Fig.\ \ref{fig:spectra}(a) for a \thRbZn\ sample at various different temperatures.

\noindent
The noise PSD of a superposition of $1/f$ noise and a single Lorentzian contribution can be written as
\begin{equation}
\frac{S_R(f)}{R^2}\cdot f=\frac{A}{f^{(\alpha-1)}}+\frac{B}{4\pi^3}\cdot \frac{f}{f^2+f_{\text{c}}^2},
\label{eq:psd}
\end{equation}
where $A$ is the magnitude of the 1/$f$-noise contribution and $\alpha$ the frequency exponent, which corresponds to the slope in a double-logarithmic plot, i.e.\ $\alpha=-\partial \ln S_V(f)/\partial \ln f$. The amplitude of the Lorentzian contribution is determined by $B=(\Delta R/R)^2 \cdot 1/(\tau_1+\tau_2)$. The corner frequency $f_{\text{c}}$, where $f  \times S_R(f)/R^2$ vs.\ $f$ exhibits a maximum, is defined by $f_{\text{c}}=1/(2\pi\tau_{\text{c}})=1/(2\pi)\cdot(1/\tau_1+1/\tau_2)$ with $\tau_1$ and $\tau_2$ describing the characteristic lifetimes of the states in a double-well potential with an energy barrier $E_{\text{a}}$ that must be overcome. For a thermally activated process it is $f_{\text{c}} = f_0' \exp{[-E_{\text{a}}/(k_{\text{B}}T)]}$ with an attempt frequency $f_0'$. 
In some cases, we found that the line width of the superimposed two-level fluctuations was broader than that of a single Lorentzian, see example shown in Fig.\ \ref{fig:spectra}(b). In this case, we employ the distributed Lorentzian model suggested in \cite{Kagawa2013} for analyzing the spectra.\\
In order to realize the charge-glass state even for \thTlZn\ with a large critical cooling rate, a heat pulse technique \cite{Hartmann2014} was employed. Thereby the sample's resistance acts as a Joule heater ($P=RI^2$) and the sample is thermally coupled to the low-temperature heat bath, see Appendix \ref{Append_Heat pulse}. Starting at low temperatures, by applying a stepwise increasing heating current, the sample can be warmed up above $T_{\text{CO}}$ in a controlled way by monitoring the two-point sample resistance. In order to prevent amplifying effects of the heating power when the sample resistance decreases abruptly at the charge-ordering transition, a constant current rather than a voltage is applied. When switching off the heating current, the sample cools down very fast with an estimated lower limit of the cooling rate of $|q|\sim 700\,$K/s near $T_{\text{CO}}$ resulting in a quenched charge state.

\section{Results and Discussion}

\subsection{\RbZn}
\subsubsection{Cooling-rate-dependent resistance}
\begin{figure}[t]
	\centering
	\includegraphics[width=1\linewidth]{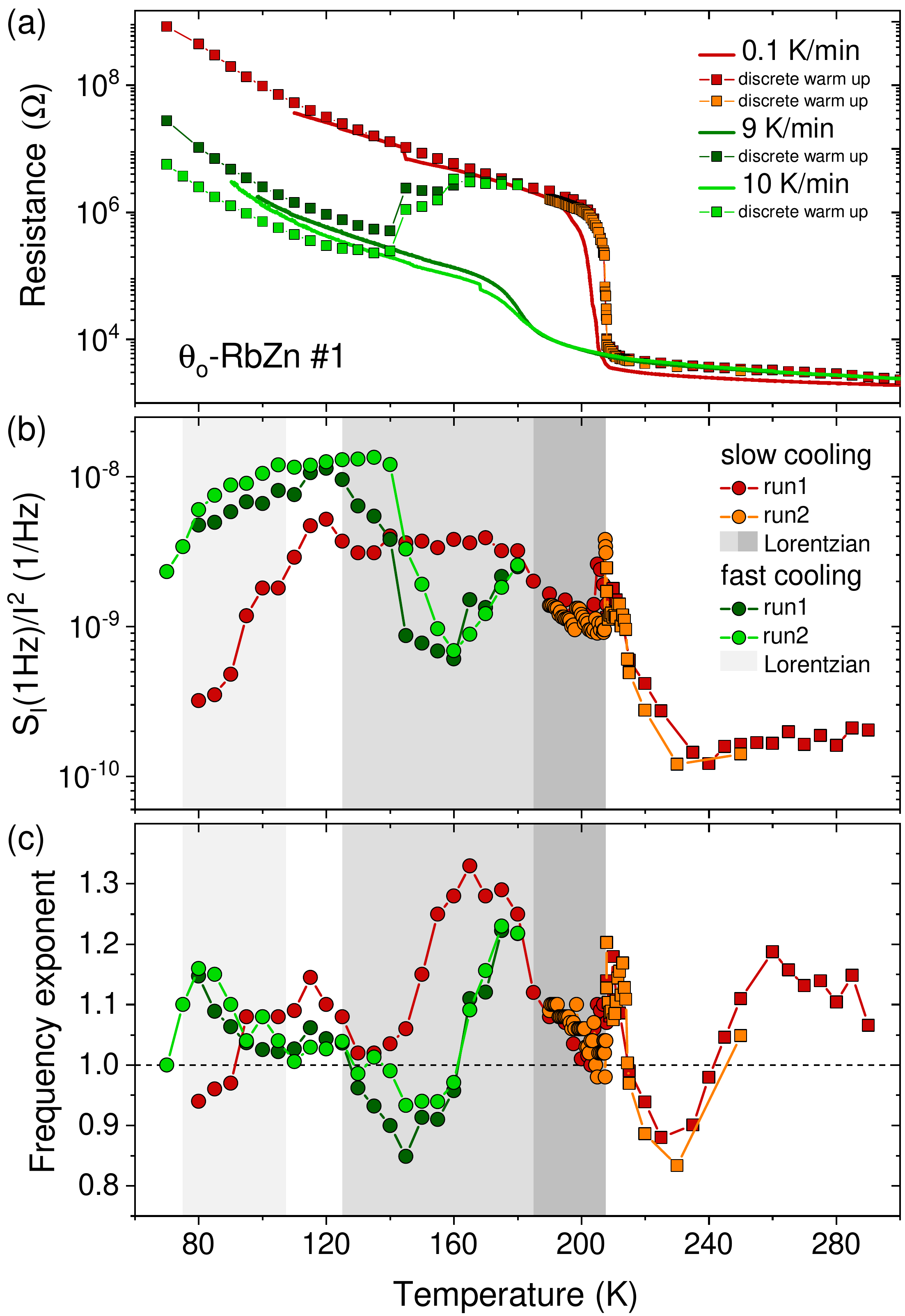}
	\caption{Resistance (a), normalized current/voltage noise PSD (b) and frequency exponent $\alpha$ (c) of the $1/f^\alpha$-noise contribution vs. temperature of \thRbZn\ (sample \#1) for the slow- and fast-cooled state. Only the first term of Eq.\,(\ref{eq:psd}) is evaluated here. The shaded areas mark the regions where additional, superimposed Lorentzian spectra emerge, see also Fig.\ \ref{fig:RbZn_Lorentz} below.}
	\label{fig:RbZn_res+noise}
\end{figure}

\begin{figure}[t]
	\centering
	\includegraphics[width=1\linewidth]{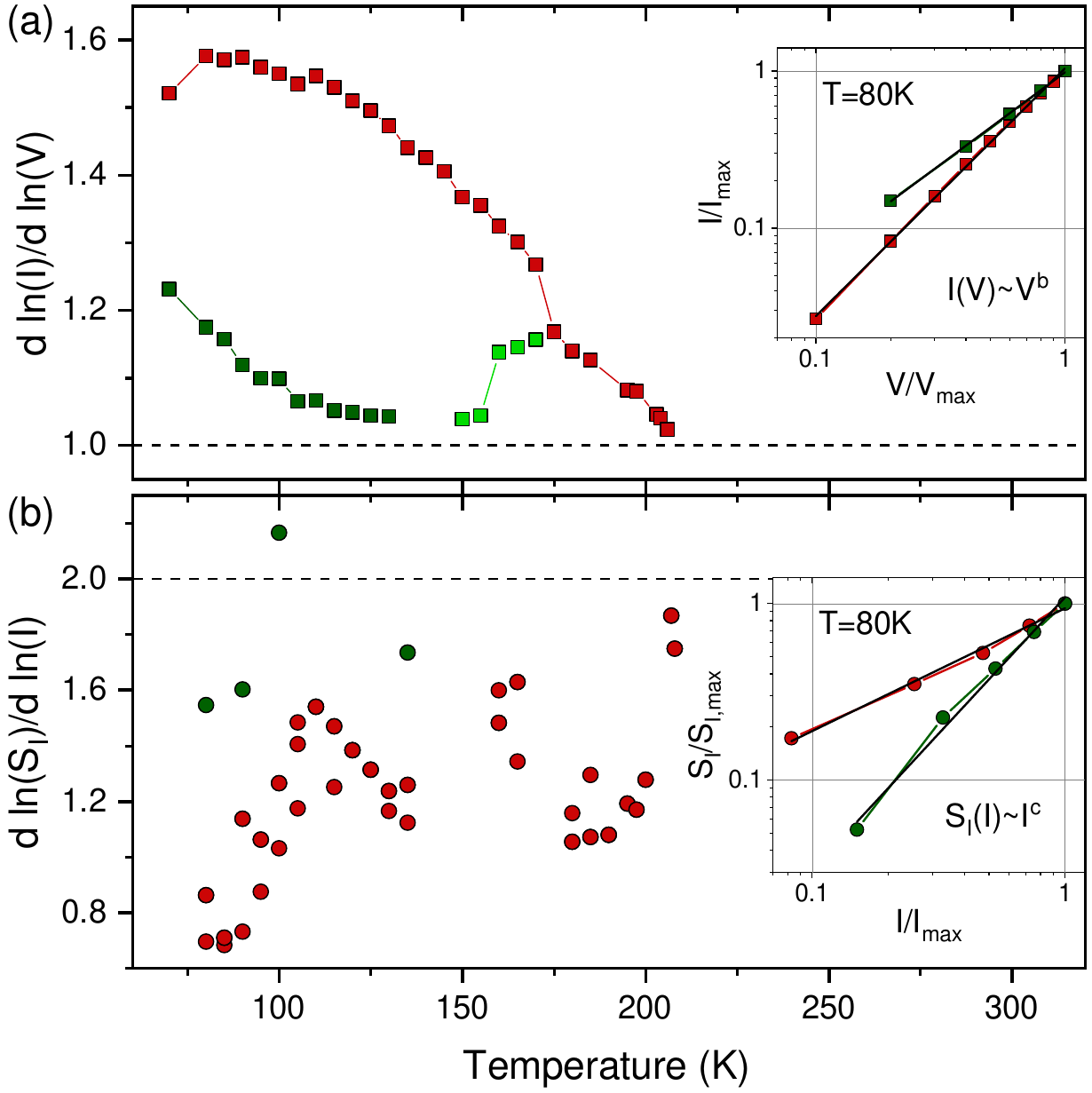}
	\caption{Power law exponents of the nonlinear $IV$ curves (a) and of the noise PSD in dependence of the current, ${\rm d}\ln(S_I)/{\rm d} \ln(I)$ (b). The dashed lines indicate ohmic behavior and quadratic current dependence of the noise, which is valid above the charge-ordering transition. Insets show the $IV$ curves (upper inset) and the current-dependent PSDs (lower inset) in a double-logarithmic plot at $T=80\,$K for the slow- (red) and fast-cooled state (green).}
	\label{fig:nonlinear}
\end{figure}

The \thRbZn\ compound with medium frustration shows a strong cooling rate dependence of the resistance, as depicted in Fig.\ \ref{fig:RbZn_res+noise}(a).
For slow cooling with $|q|=0.1\,$K/min (solid red line), a CO transition from the CL into the CC phase occurs at $T_{\text{CO}}\sim200\,$K, which is accompanied by a sudden increase of the resistance by three orders of magnitude and a clear hysteresis upon warming up, typical for a first-order phase transition. For fast cooling with $|q|\sim10\,$K/min (solid green lines) we observe a smooth continuation of the CL curve at $T_{\text{CO}}$ which means that the sharp first-order transition is suppressed and a CG state is realized. However, although the absolute resistance value is at least two orders of magnitude smaller than in the ordered phase below $T_{\text{CO}}$, we still observe a broadened, step-like increase of the resistance below the transition region at $T\sim 180\,$K, indicating that the critical cooling rate of 5\,K/min reported in the literature \cite{Kagawa2013} is too slow in our case. Analyses of the temperature-dependent resistance curves for slow and fast cooling by assuming an Arrhenius behavior at low temperatures (see Appendix \ref{Append_resistance}) and comparing the ratio of respective activation energies with the literature \cite{Takahide2006,Takahide2010} also support the hypothesis that the CO transition is not fully suppressed for the applied cooling rate. This may be due to the higher quality of the present samples, grown with an improved method, a notion that is corroborated by the observation that disorder introduced by x-ray irradiation has a strong impact on the CG forming ability \cite{Hashimoto2022}. Another possible explanation are size effects, as in \cite{Oike2018} it was reported, that a smaller sample size of the systems IrTe$_2$ and \thRbZn\ causes a higher degree of supercooling, which also affects the critical cooling rate for quenching the system.\\
For the present compound, we were able to extend previous noise measurements, which were limited to the CL state \cite{Kagawa2013,Sasaki2017}, to both the ordered CC phase below $T_{\text{CO}}$ and the metastable CG state and compare the low-frequency charge carrier dynamics in these states. The resistance and noise measurements shown in Fig.\ \ref{fig:RbZn_res+noise}(a) and (b), (c), respectively, were performed during warming up the sample in discrete temperature steps. Due to the peculiar electron (cluster) dynamics and crystallization for the fast-cooled state, at 145\,K the resistance starts to relax back to the ordered CC phase upon warming and approaches the slowly cooled curve at 180\,K. The crystallization dynamics of electrons in this temperature regime has been studied in detail \cite{Sasaki2017,Sato2017}, surprisingly revealing -- albeit the quantum nature of electrons -- the same nucleation and growth processes that characterize conventional glass formers.\\
An important issue in the CO phase are nonlinear $IV$ curves with power law exponents $b\equiv {\rm d}\ln(I)/{\rm d} \ln(V)$ up to 1.6 (red squares in Fig.\ \ref{fig:nonlinear}(a)), which were suggested to originate in electric field-induced unbinding of thermally activated electron-hole pairs  \cite{Takahide2006,Takahide2010}. These intrinsic nonlinear effects also have been discussed in \cite{Alemany2015,Nogami2010,Watanabe2008} and assigned to the current-induced melting of CO, which can develop from the avalanche-like creation of many electron-hole pairs. 
X-ray diffuse scattering experiments \cite{Nogami2010} have shown, that the ${\bf {\it q}}_2^\prime = (0,k,1/2)$ modulation, whose 3D order is associated with the metal-insulator transition, is unstable against strong electric current, resulting in a decrease of the number of domains which exhibit the local ${\bf {\it q}}_2^\prime$ order. In contrast to the CC state, the CG state (green squares in Fig.\ \ref{fig:nonlinear}(a)) shows almost linear $IV$ characteristics above about 100\,K before it starts to deviate from ohmic behavior upon further cooling, in accordance with \cite{Takahide2010}. 
Since the $IV$ curves were taken upon warming up the sample in discrete steps, the degree of nonlinearity approaches that of the slow-cooled CO state during relaxation of the quenched state at $T \sim180\,$K. The implications of these nonlinearities for the noise measurements are discussed below.

\subsubsection{Resistance noise in the slow-cooled state}
The noise measurements which were performed in a two-terminal (circular symbols) and four-terminal (squares) DC configuration agree well as demonstrated by their overlap for the temperatures where both setups have been used. A comparison of the noise results for the CC and CG states reveals important differences in the respective charge carrier dynamics (see Fig.\ \ref{fig:RbZn_res+noise}(b) and (c) for magnitude and frequency exponent, respectively). Another striking observation is a sharp and pronounced peak in both the noise magnitude and the frequency exponent at the temperature of the transition from the CO to the CL phase upon warming at $T_{\text{CO}}$, see Fig.\ \ref{fig:RbZn_res+noise}(b) and (c). Such an enhancement of the noise magnitude and strong shift of spectral weight to low frequencies ($\alpha > 1$) have been observed for another \thRbZn\ sample (not shown) and may be related to  fluctuations of microscopic entities which play a key role in phase transitions, also of first-order,  in two dimensions \cite{Chen2007}.\\
Remarkably, the normalized noise PSD of the slowly cooled state (red/orange color) becomes strongly enhanced already at temperatures $T \lesssim 230$\,K, i.e.\ about $25 - 30$\,K above the transition (cf.\ the resistance $R(T)$), as compared to the rather flat behavior in the CL phase at higher temperatures. The rise of the noise magnitude upon approaching the transition from above is a clear signature of slowly fluctuating charge clusters as charge disproportionation has been observed already in the metallic phase above $T_{\text{MI}}$ \cite{Takahashi2004,Chiba2004}. This behavior continues upon cooling through $T_{\text{CO}}$ down to $T\sim160\,$K, where a plateau develops. Along with this increase of the magnitude of the fluctuations we observe a concurrent change of the frequency exponent from $\alpha = 0.8$ at $T = 230$\,K to $\alpha =1.3$ at $160\,$K. This substantial shift of spectral weight to lower frequencies is in accordance with the freezing of the charge clusters' dynamics \cite{Kagawa2013,Sasaki2017,Sato2017}. As pointed out above, note that superimposed on this broad evolution of the charge fluctuation dynamics to slower timescales are the sharp peaks in magnitude and frequency exponent at the phase transition $T_{\text{CO}}$. 
For lower temperatures $T \lesssim 120$\,K the PSD apparently decreases again, accompanied by a frequency exponent $\alpha$ around 1, which implies a more homogeneous distribution of the relevant dynamic time scales.

\begin{figure}[t]
	\centering
	\includegraphics[width=1\linewidth]{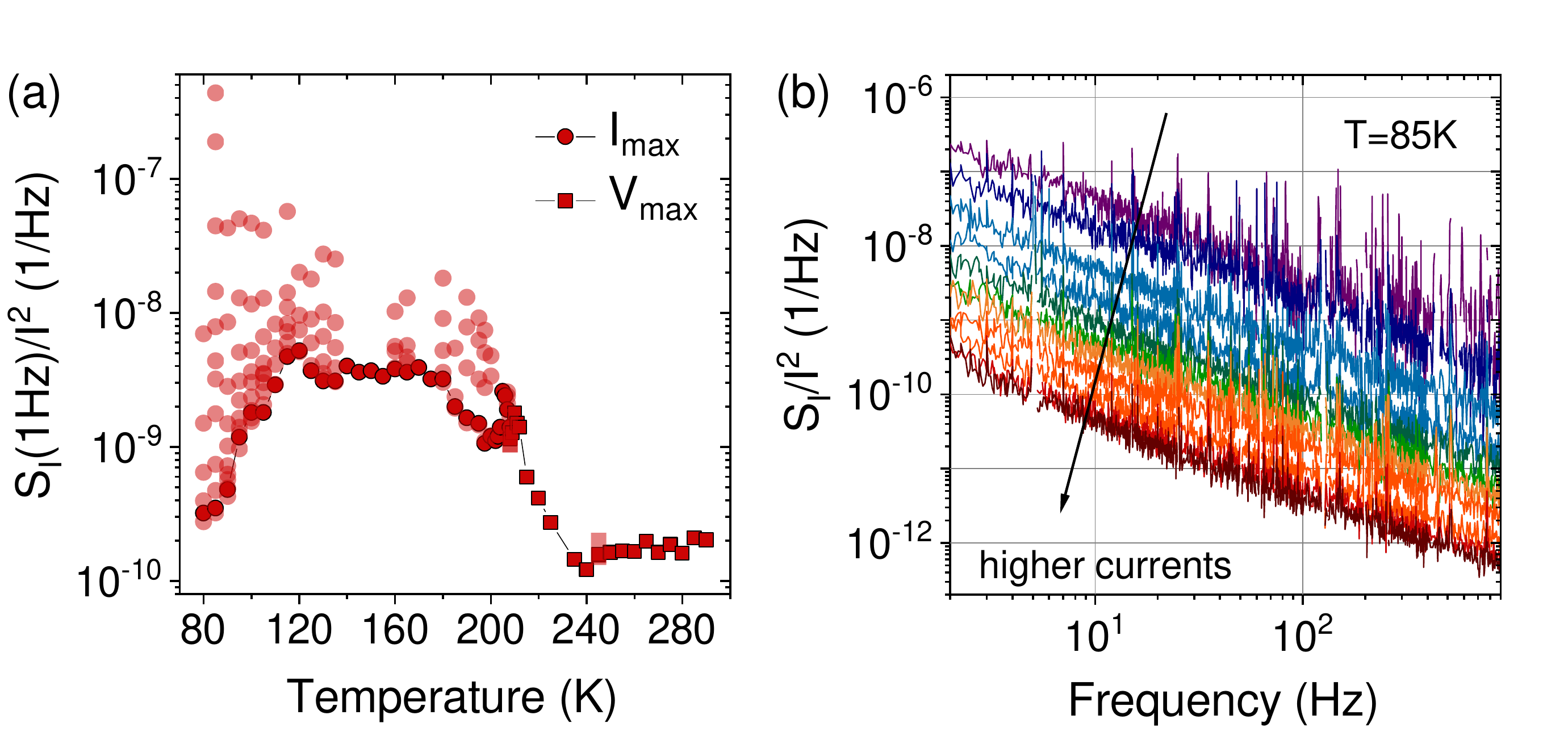}
	\caption{Normalized PSD taken at $1\,$Hz vs. temperature for different currents (a). Dark symbols indicate the values for the maximum current or voltage, transparent symbols represent the values for smaller currents/voltages. Spectra for different currents are exemplarily shown for $T=85\,$K (b).}
	\label{fig:nonlinear_spektren}
\end{figure}

\noindent At this point it is necessary to discuss the nonlinearities, described above for the resistance or conductivity, also for the PSD of the voltage or current fluctuations, which usually scale with the voltage or current squared \cite{Hooge1969,Kogan1996}, i.e.\ $S_I\propto I^2$. Interestingly, this scaling relation is not observed in the CC state, where the data can be described by $S_I \propto I^c$ with $c \sim 0.7-1.6$. The power-law exponent $c\equiv {\rm d}\ln(S_I)/{\rm d} \ln(I)$ of the current-dependent PSD for the slow- (red) and fast-cooled (green) state is illustrated in Fig.\ \ref{fig:nonlinear}(b). Since the nonquadratic noise is most pronounced in the charge-ordered phase, where also nonlinear $IV$ curves in the time-averaged transport occur, the transport mechanism responsible for this nonlinearity may also influence the strength of fluctuations. The current-dependent normalized noise PSD $S_I/I^2$ makes it now difficult to compare its magnitudes at different temperatures, since the maximum applied current varied in the experiment (in order to keep the Joule power roughly constant and thereby avoid sample heating) as well as the strength of nonquadraticity of the noise varies as a function of temperature.\\
The PSD normalized to $I^2$ at $1\,$Hz evaluated for different currents is shown in Fig.\ \ref{fig:nonlinear_spektren}(a). The value for the maximum applied current or voltage is indicated by the dark red circles and squares (cf. Fig.\ \ref{fig:RbZn_res+noise}(b)), whereas the values for lower currents or voltages are represented by transparent symbols. By comparing $S_I/I^2(1\,\text{Hz})$ for different currents and temperatures, it can be seen that the strongest variation is below $120\,$K, also indicated by the small value of the power law exponent $c$ (Fig.\ \ref{fig:nonlinear}(b)). This is visualized in Fig.\ \ref{fig:nonlinear_spektren}(b), where the normalized spectra measured at $T=85\,$K vary by almost three orders of magnitude. Therefore, we assign the strong decrease of the noise level with the maximum applied current at low temperatures ($T<120\,$K) to the nonquadraticity of the noise. The strong reduction of the noise PSD for higher currents might be caused by electric-field induced free charge carriers, also responsible for the nonlinear $IV$ characteristics \cite{Takahide2010}, resulting in an enhanced number of fluctuators and therefore a lower overall noise level due to cancellation effects. Likewise, the current-induced melting of CO domains \cite{Alemany2015,Nogami2010,Watanabe2008} may increase the number of conduction paths or the noisy volume. We note that nonquadratic noise has also been observed in other systems \cite{Parman1991,Carbone2001,Bardhan2005}, see also \cite{Kogan1996}. A mathematical treatment of this phenomenon is given in Appendix \ref{Nonquadratic noise}.

\subsubsection{Dominating two-level processes}
\begin{figure}[t]
	\centering
	\includegraphics[width=1\linewidth]{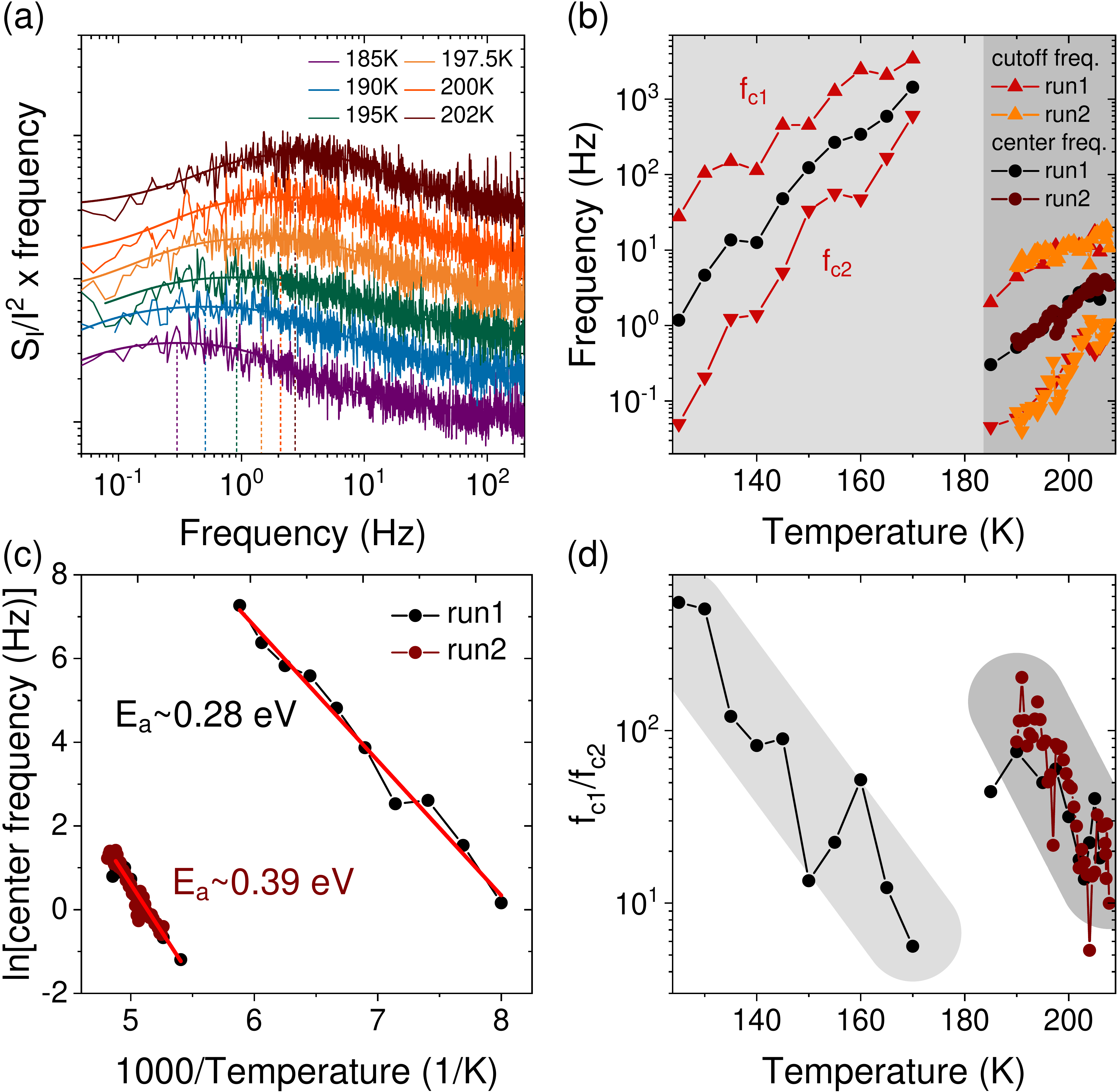}
	\caption{(a) Examples of Lorentzian contribution superimposed on the $1/f$ spectra in the temperature region 125-205\,K measured in the CC state of \thRbZn. (b) $f_{\text{c}1}$ and $f_{\text{c}2}$ stands for the high- and low-frequency cutoffs, $f_{0}=\sqrt{f_{\text{c}1}\cdot f_{\text{c}2}}$ for the center frequency. (c) Extracted energies from the center frequency and (d) line width of the distributed Lorentzian contribution.}
	\label{fig:RbZn_Lorentz}
\end{figure}

In addition to the $1/f^\alpha$ fluctuations, we find superimposed Lorentzian contributions for temperatures $125\,{\rm K} < T < T_{\rm CO}$ for slow cooling, which implies one or a few dominant fluctuating processes. Hence, it should be noted that the values for $S_I(1\,\text{Hz})/I^2$ shown in Fig.\ \ref{fig:RbZn_res+noise}(b) represent the underlying $1/f$ noise. The analysis of the non-$1/f$ contribution, fitted by a distributed Lorentzian model \cite{Kagawa2013} (solid lines in Fig.\ \ref{fig:RbZn_Lorentz}(a)), reveals two different processes dominating the noise in the temperature regions $125\,{\rm K}  \lesssim T  \lesssim185\,$K and $185\,{\rm K} \lesssim T  \lesssim T_{\rm CO}$ (gray shaded regions in Fig. \ref{fig:RbZn_res+noise}).

\noindent
From the high- and low-frequency cutoffs $f_{\text{c1}}$ and $f_{\text{c2}}$ we extract a center frequency $f_0 = \sqrt{f_{\text{c1}} f_{\text{c2}}}$ (black and brown symbols in Fig.\ \ref{fig:RbZn_Lorentz}(b)), which follows an Arrhenius law $f_0 \propto \exp{[-E_{\text{a}}/(k_{\text{B}}T)]}$, i.e.\ the frequency shifts to higher values for increasing temperatures as shown in Fig.\ \ref{fig:RbZn_Lorentz}(a). The corresponding activation energies $E_{\text{a}}$, which dominate the charge fluctuations in the CC phase, are $0.28\,$eV and $0.39\,$eV, cf.\ Fig.\ \ref{fig:RbZn_Lorentz}(c). One can speculate that these energies correspond to the energy barrier of the switching between metallic and insulating phases or between charge clusters with different local orders. From the ratio of the cutoff frequencies, shown in Fig.\ \ref{fig:RbZn_Lorentz}(d), we can see, that for both processes the line width $f_{\text{c1}}/f_{\text{c2}}$ becomes broader for decreasing temperatures implying that the dynamics becomes more heterogeneous. Similar features of slow and heterogeneous dynamics were found in the CL state for different $\theta$-$MM^\prime$ compounds \cite{Kagawa2013,Sato2014,Sato2016,Sasaki2017} and were ascribed to the freezing of charges on the triangular lattice.\\
The observed two-level processes in the slow-cooled CC state with pronounced Lorentzian contributions may be attributed to electronic phase separation/phase coexistence near the first-order charge-ordering transition \cite{Bid2003,Daptary2019}. However, such processes are also observed in the CC phase even well below the metal-insulator transition. Possibly the switching of metallic and insulating clusters or of domains with different competing local orders persists even deeper in the insulating phase. An incomplete formation of charge order would explain the existence of slow and heterogeneous dynamics of charge clusters freezing in a glassy manner even below $T_{\text{CO}}$, originating from a small liquid (metallic) phase. Another possible origin of coexisting phases is a (current-induced) destabilization of the CO state due to high electric fields, since in the slow-cooled state also nonlinear $IV$ curves were observed.

\subsubsection{Resistance noise in the partly quenched state}
The noise level of the partly quenched state (green symbols in Fig.\ \ref{fig:RbZn_res+noise}(b)) is about one order of magnitude higher at low temperatures but undershoots the slow-cooled curve at the temperature where relaxation begins. For further warming, the PSD monotonically increases and reaches the slow-cooled curve after completed charge crystallization. Notably, in the relaxation region, repeated noise measurements at a fixed temperature fall on one curve when normalizing the PSD to the actual resistance at that time. This means, that within the measuring time the fluctuating processes do not change, although the resistance value changes. In contrast, the temperature has a strong influence on the noise, which monotonically increases with increasing $T$ until the curve of the slowly cooled state is reached. A possible explanation are different nucleation mechanisms for the temperature regimes, as discussed in \cite{Sato2017}, where below the nose temperature of the time-temperature-transformation (TTT) diagram the fraction of CO domains remain undeveloped, whereas the resistance already increases with time. The coexistence of two competing local orders 
have been observed for intermediate cooling speed by x-ray diffuse scattering \cite{Watanabe2003,Nogami2010}, where heating leads to the growth of the thermodynamically stable ${\bf {\it q}}_2^\prime=(0,k,1/2)$ modulation. We therefore assign the increase of the low-frequency fluctuations in the relaxation region to the crystallization process, where charge clusters start to grow and rearrange.\\
In the temperature region $75\,{\rm K} \lesssim T \lesssim 105\,$K we observe (single) Lorentzian spectra superimposed on the $1/f$ spectra. From the corner frequency, which follows an Arrhenius law, an energy of $E_{\text{a}}\sim180\,$meV is deduced, which is significantly lower than the energies observed for the CC state. Since the fast-cooled state at low temperatures is most likely a mixture of CC and CG phases, the higher noise magnitude might be due to an inhomogeneous current flow, which results in a small noisy volume and therefore leads to a large noise magnitude.
Since the sample was cooled down with a continuous helium-flow cryostat with variable temperature insert (VTI), the maximum cooling rate in the relevant temperature range was limited to $\sim10\,$K/min. In order to compare the slow-cooled state with the fully quenched charge-glass state, a second sample of \thRbZn\ was examined with the aid of a heat pulse method.

\subsubsection{Resistance noise in the fully quenched state}

\begin{figure}[b]
	\centering
	\includegraphics[width=1\linewidth]{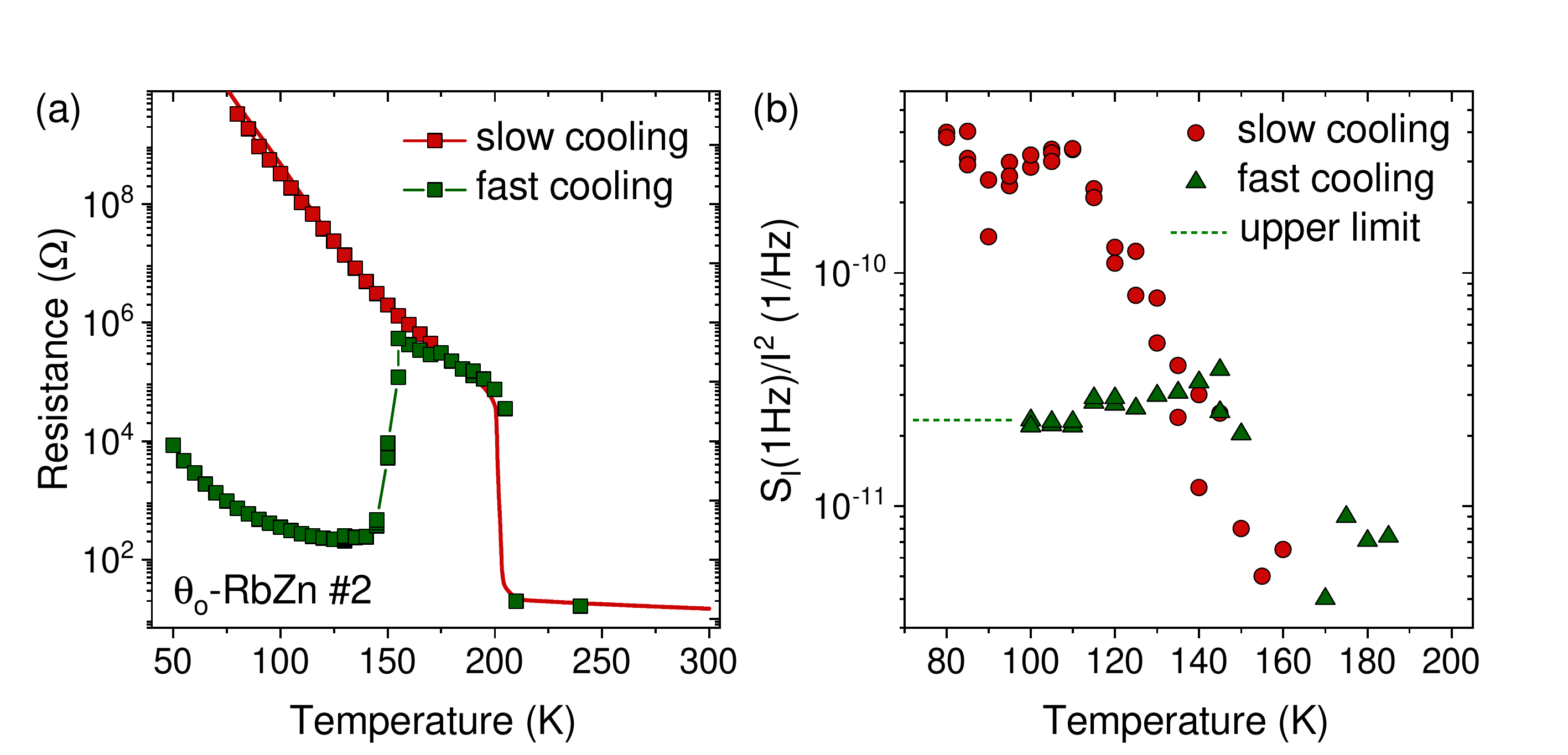}
	\caption{Resistance (a) and normalized PSD taken at $1\,$Hz (b) of \thRbZn\ (sample \#2) for the slow- and fast-cooled state, which was generated by using a heat pulse.}
	\label{fig:RbZn_HS06}
\end{figure}

Figure \ref{fig:RbZn_HS06}(a) shows the resistance of \thRbZn, sample \#2, for slow ($|q|=0.1\,$K/min) (red color) and fast cooling (green color). The resistance values (square symbols) were measured when warming up the sample in discrete temperature steps after slow and fast cooling (employing the heat pulse technique). In comparison with the first sample, here the charge-ordering transition is sharper and the total increase of the resistivity amounts to $\sim9$ orders of magnitude from room temperature to 75\,K, implying a better sample quality or the appearance of size effects \cite{Oike2018} (see above), as sample \#2 has a larger volume ($V_2\sim3\cdot10^7\,\mu \text{m}^3$) than sample \#1 ($V_1\sim3\cdot10^6\,\mu \text{m}^3$). The application of the heat pulse technique with $|q|\gtrsim700\,$K/s leads to a full suppression of the transition and to huge differences in the resistance by many orders of magnitude. This is corroborated by the ratio of activation energies of the slow- and fast-cooled states assuming an Arrhenius law (see Appendix \ref{Append_resistance}), in agreement with \cite{Takahide2010}.

\noindent
For this sample, measurements of the resistance fluctuations were only possible below the transition, probably due to the small absolute resistance at high temperatures resulting in a low overall noise magnitude. The comparison of the PSD of the slow- and fast-cooled state for $T<T_{\text{CO}}$ is shown in Fig.\ \ref{fig:RbZn_HS06}(b). Below $T\sim 130\,$K the charge-glass state shows a significantly smaller noise magnitude as compared to the CC state. At the temperature, where the resistance relaxes back to the CC phase, the noise measurements yield similar results. Below $100\,$K the resistance fluctuations of the quenched state were too small compared to the noise floor of our experimental setup at these temperatures indicated by the dashed green line giving an upper limit of the PSD. Therefore, no crossing of the curves is to be expected and the enhanced noise level in the fast-cooled state for the first sample is probably caused by a mixture of different phases because of not fast enough cooling rates. These findings are in agreement with the results on \thTlZn, which are discussed in the next section.

\subsection{\TlZn}
\subsubsection{Cooling-rate-dependent resistance}
\begin{figure}[b]
	\centering
	\includegraphics[width=1\linewidth]{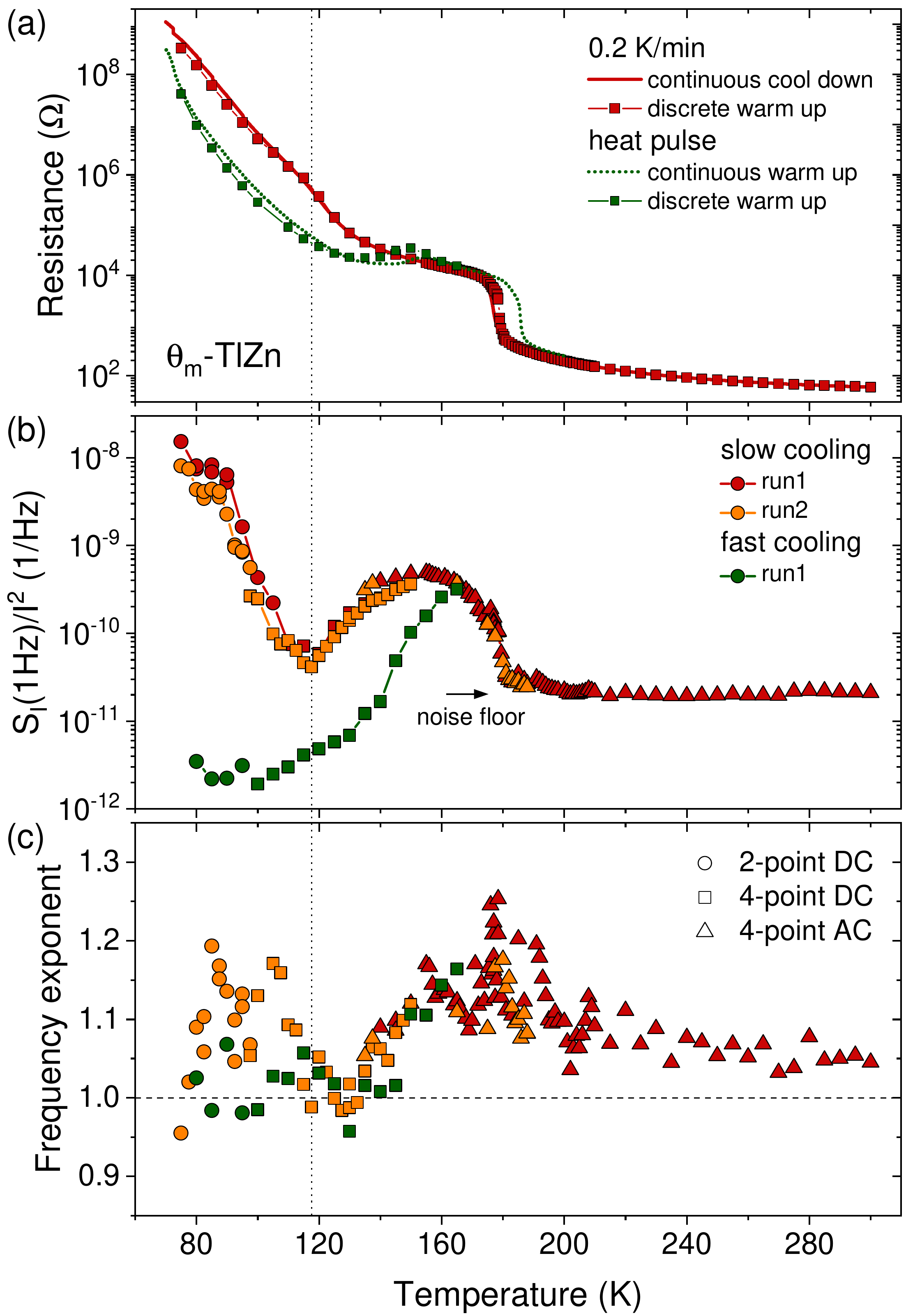}
	\caption{Resistance (a), normalized PSD (b) and frequency exponent (c) of the $1/f$ spectra measured on \thTlZn\ for the slow- (red) and fast-cooled state (green). The symbols mark different measurement techniques for fluctuation spectroscopy.}
	\label{fig:TlZn_res+noise}
\end{figure}

\thTlZn\ shows a low degree of frustration and therefore a large critical cooling rate is required to quench the charge-ordered state. The system variant with monoclinic crystal structure studied here \cite{Sasaki2017} exhibits a charge-ordering transition at $T_{\text{CO}}\approx175\,$K with clear hysteresis, see Fig.\ \ref{fig:TlZn_res+noise}(a) (red line). In the temperature region around $130\,$K the slope of the resistance curve increases for decreasing temperatures, possibly due to a change of the transport mechanism, e.g.\ a stronger localization of the charge carriers, and another resistance anomaly is seen at around 115\,K in the CC state. 

\noindent
Due to the large anisotropy of the triangular lattice in \thTlZn\ and thus a strong tendency for charge crystallization, we again used a modified heat pulse technique as described in \cite{Hartmann2014} to achieve cooling rates larger than $|q_{\text{c}}|\sim50\,$K/min required to realize the CG state. The comparison of the resistance curves for slow cooling (red curve) and for the resistance after applying a heat pulse and warming up again (green curve) is shown in Fig.\ \ref{fig:TlZn_res+noise}(a). 
Although the quenched state has been reached resulting in two distinct resistance curves of the CG and CC states, the difference only amounts to about one order of magnitude in the region $T=80 -120\,$K which is much smaller than for the \thRbZn\ compound, which may be ascribed to the different position in the phase diagram \cite{Sasaki2017}. (The steeper slope ${\rm d}R/{\rm d}T$ just above the transition indicates a more insulating behavior.)
For the quenched state, at $T\gtrsim110\,$K the resistance starts to relax back to the CC state and relaxation is completed at about $160\,$K.

\subsubsection{Resistance noise in the slow-cooled state}

Noise measurements were performed during warming up the sample in discrete temperature steps for both charge states, which revealed spectra of $1/f$-type for the whole temperature range. For temperatures $T<T_{\text{CO}}$, a two-terminal DC current noise (circular symbols) as well as a four-terminal DC voltage noise technique (square symbols) was used, whereas in the charge-liquid state a four-terminal AC voltage noise method (triangular symbols) was better suitable. The normalized noise magnitude at 1\,Hz for the slowly cooled state (red and orange symbols in Fig.\ \ref{fig:TlZn_res+noise}(b)) is almost constant at high temperatures and reveals a pronounced increase starting just around 175\,K, which coincides with the charge-ordering temperature. One should note that the noise level above the transition is mainly limited by the noise floor of the experimental setup in this temperature region (marked by an arrow). 
Right at the first-order transition, again the frequency exponent is enhanced showing a peak structure similar to the behavior of \thRbZn\ (Fig.\ \ref{fig:RbZn_res+noise}). Below $T_{\text{CO}}$, in the CC phase the noise magnitude shows a broad maximum around $T\sim150\,$K, which concurs with the flat (plateau-like) resistance curve. This may indicate that a not yet fully completed charge order causes the strong resistance fluctuations. Below $T\sim120\,$K, coinciding with the resistance anomaly (marked by the dotted line), the noise PSD shows a sharp increase by almost three orders of magnitude upon further lowering the temperature. (The latter can be described equally well with an exponential function and a power law behavior which implies a scaling $S_I/I^2\sim R^{\nu}$ expected for a percolative system \cite{JMueller2009a}; here, we find $\nu = 0.94$.) The frequency exponent of the $1/f$ spectra (Fig.\ \ref{fig:TlZn_res+noise}(c)) shows enhanced values up to $\alpha\sim1.3$ in the CO transition region and decreases as the noise magnitude approaches its minimum at $T\sim120\,$K. The strong increase of the noise level at low temperatures then is accompanied by larger values for $\alpha$, although the curve progression varies between 1.0 and 1.2.

\subsubsection{Resistance noise in the quenched state}

The normalized PSD at 1\,Hz for the quenched state (green symbols in Fig.\ \ref{fig:TlZn_res+noise}(b)) is four orders of magnitude lower at the lowest measured temperature ($T=75\,$K) and increases gradually upon warming up the sample until it matches the noise level of the CC state after completed relaxation. Remarkably, in contrast to the resistance, which shows only small changes when comparing the relaxed and the quenched state, the resistance noise differs drastically, such that the less ordered, glassy electronic state is much more ``quiet'' regarding its low-frequency fluctuations. The frequency exponent is almost constant with $\alpha\approx1$ for temperatures $T<120\,$K and increases up to 1.2 in the region of charge relaxation, where it coincides with the slowly cooled state. The resistance as well as the noise results have been reproduced for repeated measurements, which confirms that the sample was not modified by the large heating current and that switching between different charge states is reversible (also regarding the fluctuation dynamics), as demonstrated in \cite{Oike2015} by exploiting this effect as phase-change memory function.

\section{Conclusion}
We have investigated the low-frequency charge dynamics of different charge states in the organic conductors $\theta$-(ET)$_2$$M$Zn(SCN)$_4$ with $M=$Rb,Tl, which exhibit a varying strength of geometric frustration and electron-phonon coupling. We were able to compare the resistance/conductance noise PSDs in the CL state at high temperatures with those in both the CC and CG states at low temperatures. The compounds \thRbZn\ and \thTlZn, which exhibit a first-order transition at $T_{\text{CO}}\sim200\,$K and $\sim175\,$K, respectively, show a strong increase of the resistance fluctuations when approaching the charge-ordering temperature from above for slow cooling. Right at the transition the noise magnitude as well as the frequency exponent show pronounced peaks indicating slow dynamics due to fluctuating microscopic entities being a signature of the first-order phase transition. In the CC state of \thRbZn\ Lorentzian spectra superimposed on the $1/f$ background reveal slow and heterogeneous dynamics with characteristic energies of $0.28\,$eV and $0.39\,$eV. Accompanying nonlinear $IV$ curves we observe nonquadratic noise, i.e.\ deviations from the usual quadratic current or voltage scaling of the PSD. 
A possible scenario is that higher electric fields induce additional free charge carriers or decrease the number of CO domains, which leads to a lower noise magnitude due to an increasing number of fluctuators or noisy volume. An open question is, if the current-dependent noise is partly caused by electronic ferroelectricity, which is reported to be induced by CO in combination with a small dimerization of the ET molecules \cite{Nad2006a,Tomic2015}. A strong dependence of the PSD on the electric field was also observed in \cite{JMueller2020} and assigned to fluctuating polar clusters. The higher noise level of the fast-cooled state in \thRbZn\ is probably caused by a mixture of CC and CG phases, since measurements on a second sample, where a heat pulse technique was used to fully suppress the CO transition, reveal a lower noise magnitude for the whole temperature range. This is in agreement with results on \thTlZn\ where the CG state shows values of the PSD, which are several orders of magnitude smaller compared to those of the CC state. Possible explanations for this are strongly inhomogeneous current paths in the insulating CC state, which reduces the noisy volume and therefore leads to an increase in the noise magnitude, or much smaller energy scales of the dominating fluctuators in the CG state compared to the CC state meaning that the low-frequency fluctuations are dominant at lower temperatures. It is worth noting that a model of non-exponential kinetics by Dutta, Dimon and Horn (DDH model) \cite{Dutta1979}, based on independent thermally activated fluctuators distributed in energy, describes the temperature- and frequency-dependent $1/f^\alpha$ noise very well in highly-frustrated $\theta_o$-CsCo \cite{Thomas2022} and $\theta_o$-CsZn \cite{Sato2016}. However, it is not applicable for the samples discussed here exhibiting a charge-ordering transition. This may be interpreted in a sense that the charge dynamics in the CC state is rather governed by interacting clusters. Since the CG state for the less frustrated compounds show a different noise behavior than the one observed in $\theta_o$-CsZn and $\theta_o$-CsCo, which exhibits a structural glass transition \cite{Thomas2022}, thermal expansion measurements are highly desirable to study the influence of structural degrees of freedom on the novel charge-glass state.

\begin{acknowledgements}
We acknowledge support by the Deutsche Forschungsgemeinschaft (DFG, German Research Foundation) through TRR 288 - 422213477 (project B02).
This work was also supported by Grants-in-Aid for Scientific Research (KAKENHI) from MEXT, Japan (No. JP21H01793, JP20H05144, JP19H01833, and JP18KK0375), and Grant-in-Aid for Scientific Research for Transformative Research Areas (A) “Condensed Conjugation” (No. JP20H05869, JP21H05471) from Japan Society for the Promotion of Science (JSPS).\\
We would like to acknowledge R. Murata for synthesizing single crystals of $\theta$-RbZn.
\end{acknowledgements}

\section{Appendix}
\subsection{Crystal structure}\label{Crystal_structure}

\begin{figure}[h]
	\centering
	\includegraphics[width=1\linewidth]{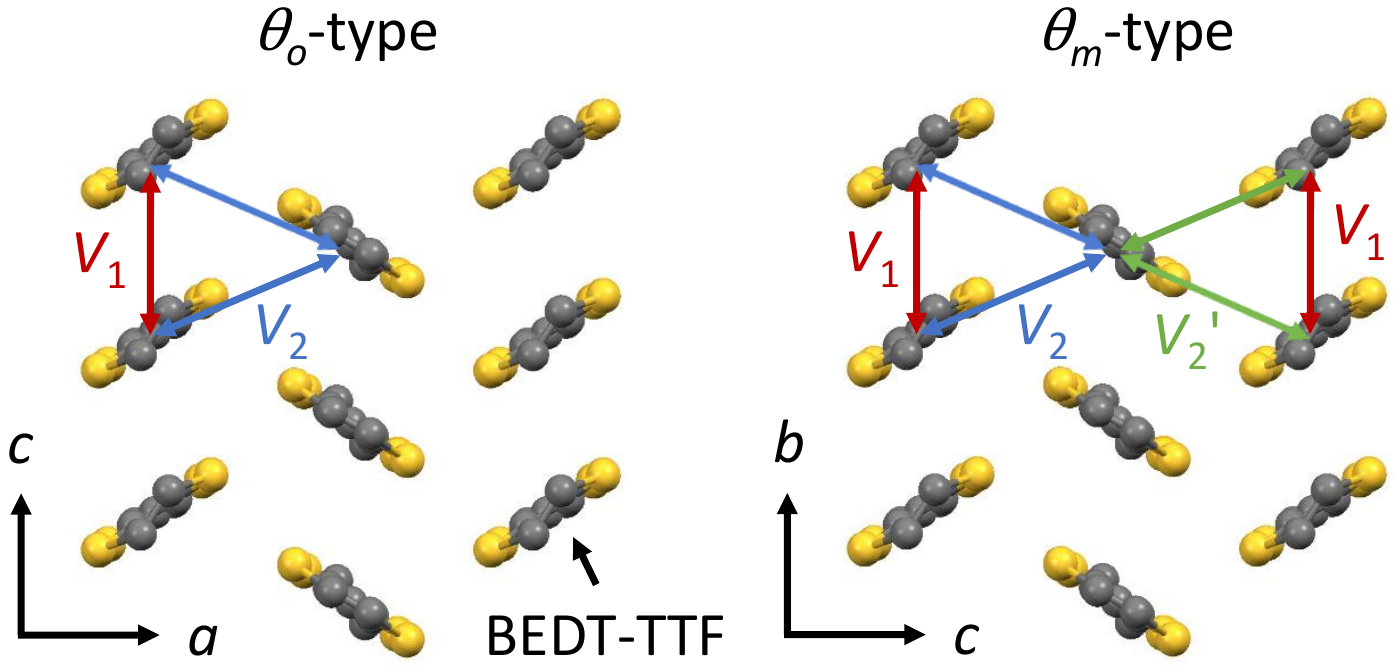}
	\caption{Molecular arrangement of the orthorhombic (left) and monoclinic (right) $\theta$-(ET)$_2MM'$(SCN)$_4$ within the conducting layers viewed along different crystallographic axes.}
	\label{fig:structure}
\end{figure}

The molecular arrangement of the donor molecules in $\theta$-(ET)$_2MM'$(SCN)$_4$ with orthorhombic and monoclinic symmetries are shown in Fig. \ref{fig:structure}. The conducting layer within the $a$-$c$ (left) and $b$-$c$ plane (right) containing the BEDT-TTF molecules can be mapped on a triangular lattice with relevant inter-site Coulomb repulsions $V_1$ and $V_2$. 
The degree of frustration determined by the anisotropy of the triangular lattice, $V_2/V_1$, is higher for $\theta_o$-RbZn ($V_2/V_1=0.87$) \cite{Sato2014a} compared to $\theta_m$-TlZn ($V_2/V_1=0.8$) \cite{Sasaki2017}.

\subsection{Heat pulse technique}\label{Append_Heat pulse}
\begin{figure}[h]
	\centering
	\includegraphics[width=0.3\linewidth]{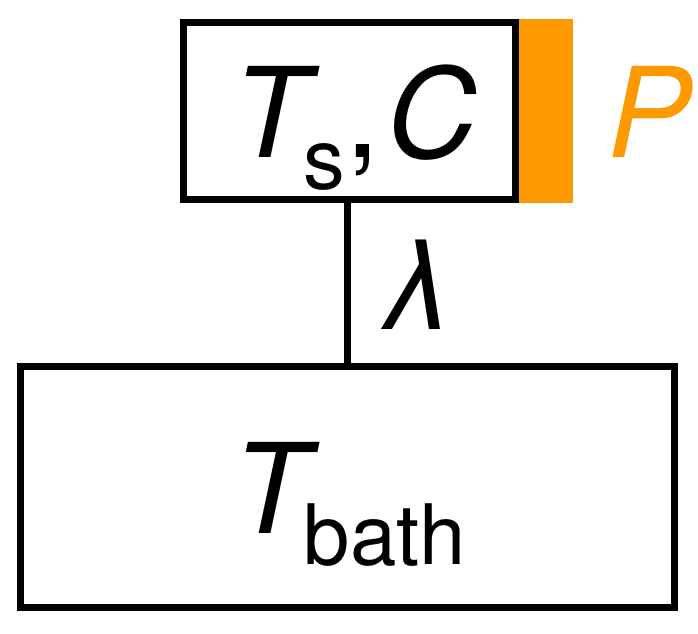}
	\caption{Simple model for the thermal conductance in the heat pulse technique after \cite{Hartmann2014}. The cold heat bath with constant temperature $T_{\text{bath}}$ is indicated by the lower rectangle, whereas the sample with temperature $T_{\text{s}}$ and specific heat $C$ corresponds to the smaller rectangle. The Joule heat $P$ (orange) is deposited by a large current flowing through the sample with resistance $R$.}
	\label{fig:heatpulse}
\end{figure}

To achieve large cooling rates in order to avoid the charge-ordering transition, a modified version of the heat pulse technique as described in \cite{Hartmann2014} is used. In this method, the sample with resistance $R$ acts as the heater itself, whereas the temperature of the bath (i.e.\ the massive copper block of the sample holder in the cryostat) stays constant. By applying a stepwise increasing current, here supplied by a sourcemeter K2400, the Joule heat $P=R I^2$ is deposited in the sample, which heats up until it reaches the equilibrium temperature $T_{\infty}$ determined by the coupling to the low-temperature heat bath. The reason for applying a constant current instead of a voltage lies in the negative resistance coefficient ${\rm d}R/{\rm d}T < 0$ at the transition, so that amplifying effects of the heating power can be avoided when the resistance always decreases during warming. By gradually increasing the current, the sample can be warmed up above the CO transition. It was always ensured that the equilibrium temperature is reached by waiting a few seconds before setting the higher current and monitoring the sample resistance by a voltmeter. When switching off the sourcemeter, the sample temperature relaxes back to the bath temperature with very high cooling rates under sufficient thermal coupling. The system can be described by a simple model \cite{Hartmann2014}, which is sketched in Fig. \ref{fig:heatpulse}.

\noindent
The sample with temperature $T_{\text{s}}$ and specific heat $C$ is thermally coupled to the heat bath with temperature $T_{\text{bath}}$ through the thermal conductivity $\lambda$. The corresponding equation for the power balance is given by
\[
P=C\cdot \dot{T_{\text{s}}}+\lambda(T_{\text{s}}-T_{\text{bath}}).
\]
The solution of this differential equation is
\[
T_{\text{s}}(t)=(T_{\text{s}}(t=0)-T_{\infty})\exp\left(-\frac{\lambda}{C}t\right)+T_{\infty}
\]

\noindent with the equilibrium temperature $T_{\infty}=\lim\limits_{t \to \infty}T_{\text{s}}(t)=\frac{P}{\lambda}+T_{\text{bath}}$. In order to obtain the temperature decay of the sample during fast cooling, we used a second sourcemeter K2400 to superimpose a small measuring current which does not lead to further warming of the sample. The fast resistance change after switching off the heating current can be determined by measuring the voltage drop with time using the buffer mode of the device. From the temperature-dependent resistance after warming up the quenched state, one in principle can extract $T(t)$ during fast cooling by matching $R(t)$ and $R(T)$. However, due to relaxation effects when warming up, this reference curve is missing in the transition region. From an underestimation of the temperature change $\Delta T$ and an overestimation of the time difference $\Delta t$, we find a cooling rate of $|q|=\left|\frac{\Delta T}{\Delta t}\right|\sim700\,$K/s as a lower limit.

\subsection{Temperature dependence of the resistance}\label{Append_resistance}

\begin{figure}[h]
	\centering
	\includegraphics[width=0.8\linewidth]{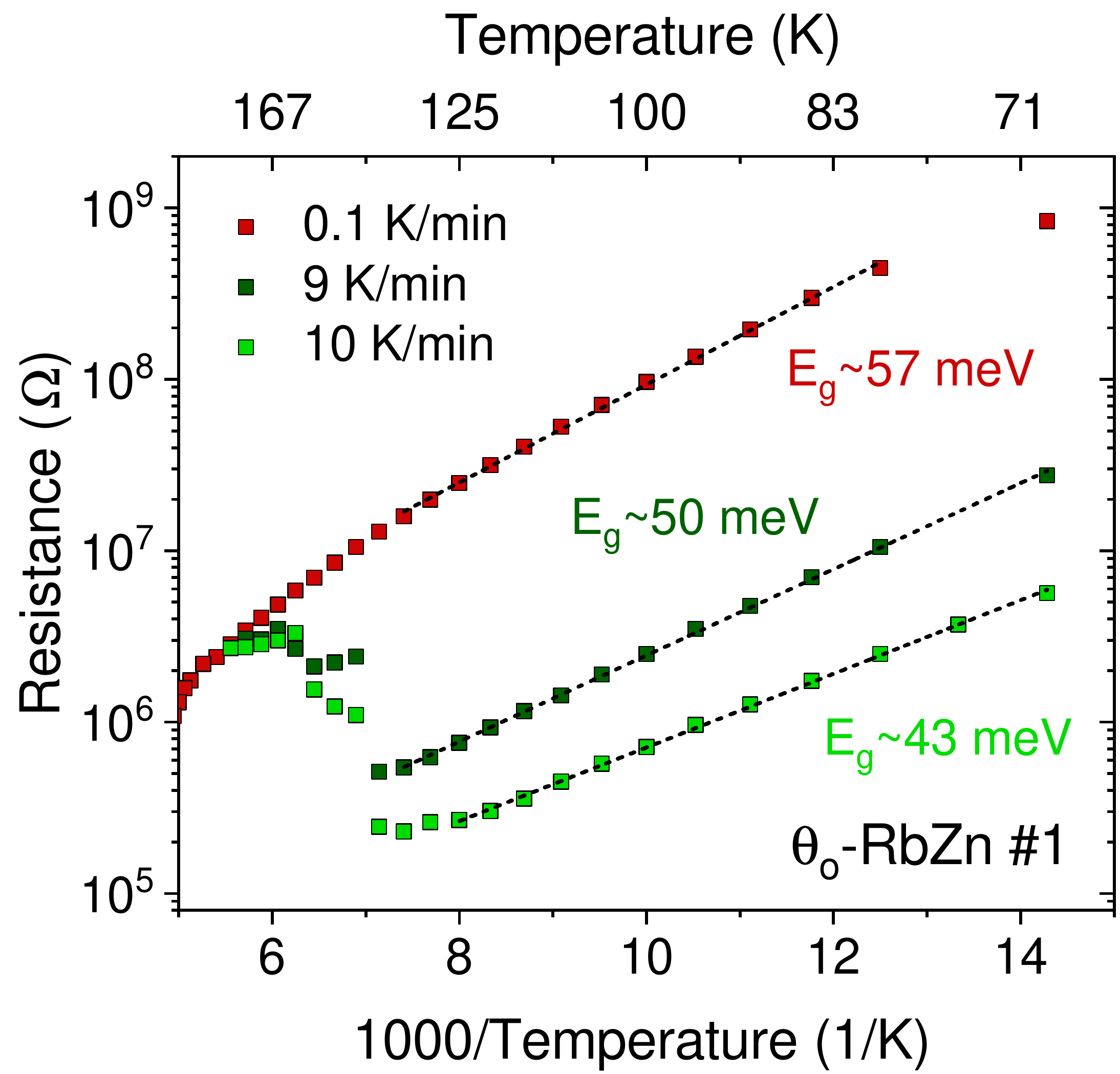}
	\caption{Logarithmic resistance of \thRbZn\ (sample \#1) against the inverse temperature of the slow- (red) and fast-cooled (green) state.}
	\label{fig:RbZn01_Arrhenius}
\end{figure}

\begin{figure}[h]
	\centering
	\includegraphics[width=0.8\linewidth]{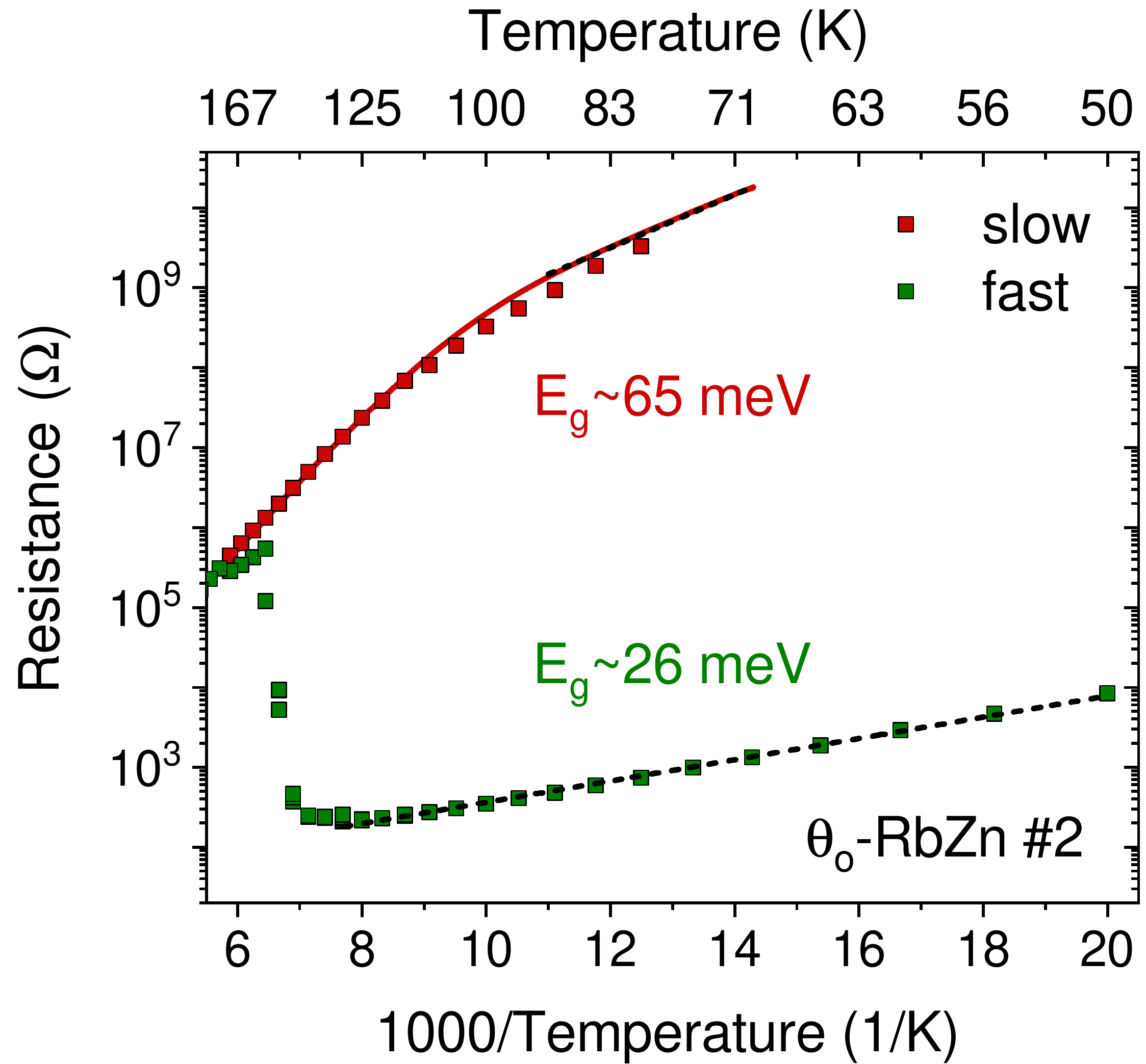}
	\caption{Logarithmic resistance of \thRbZn\ (sample \#2) against the inverse temperature of the slow- (red) and fast-cooled (green) state after the application of a heat pulse.}
	\label{fig:RbZn02_Arrhenius}
\end{figure}

The temperature-dependent resistance of the \thRbZn\ compound shows thermally activated behavior at low temperatures, i.e. $R=R_0\exp[E_{\text{g}}/(k_{\text{B}} T)]$ with an energy gap $E_{\text{g}}$, as reported previously in the literature \cite{Takahide2006,Takahide2010} (please note the factor of 2 in their definition, i.e. $E_{\text{g}}=\Delta_0/2$). The resistance was measured with a relatively low current or voltage to ensure almost linear (Ohmic) behavior. The logarithmic resistance of \thRbZn\ against the inverse temperature is shown in Fig. \ref{fig:RbZn01_Arrhenius} (sample \#1) and Fig. \ref{fig:RbZn02_Arrhenius} (sample \#2) for the slow- (red) and fast-cooled (green) state.

\noindent
The extracted energy gap at low temperatures of sample \#1 yields $E_{\text{g}} = 57\,$meV for the CO state. The energy gap for fast cooling should be much smaller according to \cite{Takahide2006,Takahide2010}, where $E_{\text{g,slow}}/E_{\text{g,fast}}\sim2.46$. Here, however, we find an only slightly smaller value of $43-50$\,meV, which indicates that the CO transition is not fully suppressed. The application of the heat pulse technique on sample \#2 leads to an activation energy of $E_{\text{g}} = 26\,$meV, matching the value found in \cite{Inada2009}. Thus the ratio of activation energies of the slow- and fast-cooled state is $E_{\text{g,slow}}/E_{\text{g,fast}}\sim2.47$ (see Fig. \ref{fig:RbZn02_Arrhenius}) in agreement with \cite{Takahide2010}, indicating the complete formation of the CG state.

\subsection{Nonquadratic noise}\label{Nonquadratic noise}	

To study the effects of nonlinear $IV$ curves on the scaling relation of the noise, we consider the Hooge law \cite{Hooge1969}, which is given by
\[
S_I=\frac{\gamma_{\text{H}} I^c}{n \Omega f^\alpha},\ \text{with\ } c=2,
\]
where $\gamma_{\text{H}} \equiv \gamma$ is an empirical parameter (dimensionless for $\alpha=1$). The current noise PSD under consideration of nonlinear $IV$ curves with $I(V) \propto V^b$ is
\[
S_I=\frac{\gamma I^2}{n \Omega f^\alpha} \propto \frac{\gamma (V^b)^2}{n \Omega f^\alpha}.
\]
By inserting the value of the power-law exponent $b = 1.6$ for the slow-cooled state in \thRbZn\ (sample \#1) at $80\,$K (cf. Fig. \ref{fig:nonlinear}(a)), one gets $3.2$ for the voltage exponent, which does not match with the experimental findings of $c<2$, cf. Fig. \ref{fig:nonlinear}(b).

\begin{figure}[h]
	\centering
	\includegraphics[width=1\linewidth]{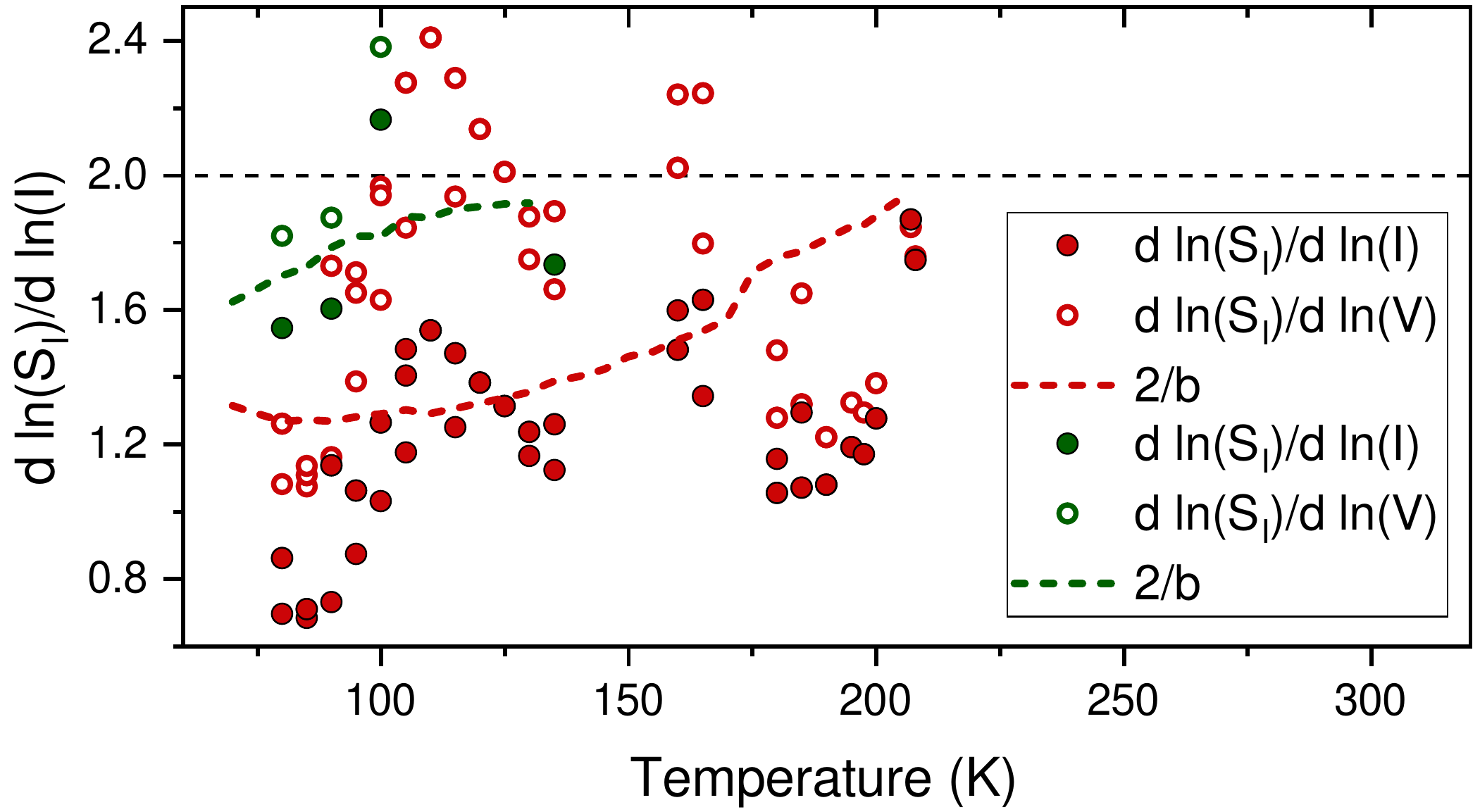}
	\caption{Power law exponents of the PSD in dependence of the current, $d\ln(S_I)/d \ln(I)$, (filled circles) as well as in dependence of the voltage, $d\ln(S_I)/d \ln(V)$, (empty circles) for the slow- (red) and fast-cooled state (green). The red and green dashed lines are the calculates values according to $c=2/b$ assuming the Hooge law under consideration of nonlinear $IV$ curves.}
	\label{fig:nonlinear_noise}
\end{figure}
\noindent
Under the assumption that the Hooge law is valid for a constant applied voltage and that $V(I)\propto I^{1/b}$, one gets
\[
S_I=\frac{\gamma V^2}{n \Omega f^\alpha} \propto \frac{\gamma I^{2/b}}{n \Omega f^\alpha}.
\]
\noindent Thus, for $T=80\,$K the current dependence of the power spectral density is expected to be $S_I\propto I^{1.25}$. The calculated exponent $c$ from the nonlinear $IV$ curves according to $c=2/b$ is represented by the red dashed line in Fig. \ref{fig:nonlinear_noise}. Although these values fit the data better, the normalization to $V^2$ should still be valid, which is not the case. It was found that $d \ln S_I/d \ln I$ (filled circles) as well as $d \ln S_I/d \ln V$ (empty circles) show deviations from 2. So another possible explanation for the deviations from Hooge's law is that the effect of nonlinear $IV$ curves, i.e. voltage-induced unbinding of thermally excited eletron-hole pairs \cite{Takahide2010}, also influences the charge carrier density $n$. Thus higher voltages lead to an increase in the number of free carriers or changes of the mobilty/number of conduction paths due to the current-induced melting of CO domains \cite{Alemany2015,Nogami2010}.
The increase of $n$ or the noisy volume $\Omega$ with the applied electric field then compensates the increase of $S_I$ with the current squared so that the growth is weaker than usual which leads to a decrease in the normalized noise magnitude.
To quantitatively analyze the influence of nonlinear $IV$ curves on the resistance fluctuations, we consider the current-dependent resistance $R(I)$ analogous to \cite{Bardhan2005},
\begin{eqnarray*}
	R(I)=R_0 \cdot f(I,c_1,c_2),
\end{eqnarray*}
where $R_0$ is the resistance for no applied current and $f$ is a function which satisfies $f(0,c_1,c_2)=1$. The resistance fluctuations $\langle(dR)^2\rangle/{R^2}$ can be determined according to:

\begin{widetext}
	\begin{eqnarray*}
		(dR)^2&=&[(dR_0)f+R_0df]^2=(dR_0)^2f^2+R_0^2(df)^2+2(dR_0)fR_0 df \\
		\frac{(dR)^2}{R^2}&=&\frac{(dR_0)^2}{R_0^2}+\frac{(df)^2}{f^2}+2\frac{df}{f}\frac{dR_0}{R_0}\\
		df&=&\frac{\partial f}{\partial I} dI+\frac{\partial f}{\partial c_1} dc_1+\frac{\partial f}{\partial c_2} dc_2\\
		(df)^2&=&\left(\frac{\partial f}{\partial I}\right)^2(dI)^2+\left(\frac{\partial f}{\partial c_1}\right)^2(dc_1)^2+\left(\frac{\partial f}{\partial c_2}\right)^2(dc_2)^2 + 2\left[\frac{\partial f}{\partial I}dI\frac{\partial f}{\partial c_1}dc_1+\frac{\partial f}{\partial I}dI\frac{\partial f}{\partial c_2}dc_2 +\frac{\partial f}{\partial c_1}dc_1\frac{\partial f}{\partial c_2}dc_2\right]\\
		\frac{(dR)^2}{R^2}&=&\frac{(dR_0)^2}{R_0^2}+{\left(\frac{\partial \ln f}{\partial I}\right)^2(dI)^2+\sum_i\left(\frac{\partial \ln f}{\partial c_i}\right)^2(dc_i)^2} + 2\sum_i \frac{\partial \ln f}{\partial I}\frac{\partial \ln f}{\partial c_i}dIdc_i+\sum_{i\neq j}\frac{\partial \ln f}{\partial c_i}\frac{\partial \ln f}{\partial c_j}dc_idc_j\\
		&+&2\frac{dR_0}{R_0} {\left(\frac{\partial \ln f}{\partial I} dI+\sum_i\frac{\partial \ln f}{\partial c_i} dc_i\right)}
	\end{eqnarray*}
	Time avering and assuming an ideal current source, i.e. $\langle(dI)^2\rangle=0$, $\langle dIdc_i\rangle=0$ and $\langle dR_0dI\rangle=0$, yields:
	\begin{eqnarray*}
		\frac{\langle(dR)^2\rangle}{R^2}&=&\frac{\langle(dR_0)^2\rangle}{R_0^2}+{\sum_i\left(\frac{\partial \ln f}{\partial c_i}\right)^2\langle(dc_i)^2\rangle}{+\sum_{i\neq j}\frac{\partial \ln f}{\partial c_i}\frac{\partial \ln f}{\partial c_j}\langle dc_idc_j\rangle}+\frac{2}{R_0} \sum_i\frac{\partial \ln f}{\partial c_i} \langle dR_0dc_i \rangle
	\end{eqnarray*}
	\text{For a specific function $f(I,K,b)=(1+K\cdot I^{b})$, which is used to fit the nonlinear $IV$ curves, we get:}\\
	\begin{eqnarray*}
		\frac{\langle(dR)^2\rangle}{R^2}&=&\frac{\langle(dR_0)^2\rangle}{R_0^2}+\left(\frac{\partial \ln f}{\partial K}\right)^2\langle(dK)^2\rangle+\left(\frac{\partial \ln f}{\partial b}\right)^2\langle(db)^2\rangle +2\frac{\partial \ln f}{\partial K}\frac{\partial \ln f}{\partial b}\langle dKdb\rangle\\
		&+&\frac{2}{R_0} \left(\frac{\partial \ln f}{\partial K} \langle dR_0dK\rangle+\frac{\partial \ln f}{\partial b} \langle dR_0db\rangle \right)\\
		\frac{\langle(dR)^2\rangle}{R^2}&=&\frac{\langle(dR_0)^2\rangle}{R_0^2}+\frac{I^{2b}}{(KI^b+1)^2}\langle(dK)^2\rangle+\frac{K^2I^{2b}(\ln(I))^2}{(KI^b+1)^2}\langle(db)^2\rangle+2\frac{KI^{2b}\ln(I)}{(KI^b+1)^2}\langle dKdb\rangle \\
		&+&\frac{2}{R_0}\left(\frac{I^b}{KI^b+1} \langle dR_0dK\rangle+\frac{KI^b\ln(I)}{KI^b+1} \langle dR_0db\rangle \right)\\
		\frac{\langle(dR)^2\rangle}{R^2}&=&\frac{\langle(dR_0)^2\rangle}{R_0^2}+\frac{1}{(K+I^{-b})^2}\left[\langle(dK)^2\rangle+{K^2(\ln(I))^2}\langle(db)\rangle^2+2K\ln(I)\langle dKdb\rangle\right]\\
		&+&\frac{2}{R_0(K+I^{-b})} \left[\langle dR_0dK\rangle+K\ln(I) \langle dR_0db\rangle \right]
	\end{eqnarray*}
\end{widetext}

\noindent From this we can see, that the normalized PSD in the Ohmic state, $\langle(dR_0)^2\rangle/R_0^2$, gets only reduced by a negative contribution of the mixed terms in the brackets. This means, there must be a correlation between the parameters $dK$ and $db$ and/or between $dR_0$ and $dK$/$db$.

\bibliographystyle{apsrev4-2}

\end{document}